\documentclass[twocolumn,showpacs,superscriptaddress,10pt]{revtex4-1}
\usepackage{graphicx,color,amssymb,amsthm,hyperref,amsmath}

\begin{document}

\def\<{\langle}
\def\>{\rangle}
\newcommand{\bra}[1]{\langle {#1} |}
\newcommand{\ket}[1]{| {#1} \rangle}
\def\be{\begin{equation}}
\def\ee{\end{equation}}
\def\bea{\begin{eqnarray}}
\def\eea{\end{eqnarray}}
\def\qed{\leavevmode\unskip\penalty9999 \hbox{}\nobreak\hfill
     \quad$\blacksquare$
     \par\vskip3pt}
\def\Pr{\textrm{Pr}}
\def\h{\frac{1}{2}}
\def\hh{\frac{1}{2}}
\newtheorem{thm}{Theorem}
\newtheorem{conjecture}{Conjecture}

\title{How Einstein and/or Schr\"odinger should have discovered Bell's Theorem in 1936}
\author{Sania Jevtic}
\affiliation{Mathematical Sciences, John Crank 501, Brunel University, Uxbridge UB8 3PH, United Kingdom}
\author{Terry Rudolph}
\affiliation{Department of Physics, Imperial College London, London SW7 2AZ, United Kingdom}

\begin{abstract}
We show how one can be led from considerations of quantum steering to Bell's theorem. We begin with Einstein's demonstration that, assuming local realism, quantum states must be in a many-to-one (``incomplete'') relationship with the real physical states of the system. We then consider some simple constraints that local realism imposes on any such incomplete model of physical reality, and show they are not satisfiable. In particular, we present a very simple demonstration for the absence of a local hidden variable incomplete description of nature by steering to two ensembles, one of which contains a pair of non-orthogonal states. Historically this is not how Bell's theorem arose - there are slight and subtle differences in the arguments - but it could have been.
\end{abstract}
\maketitle

In this paper we attempt a little revisionist history. In
particular, we show how a very simple argument
establishing the impossibility of a local realistic
description of quantum theory - Bell's theorem - was lingering on the edge of Schr\"odinger's and
Einstein's consciousness in 1935-36.

\section{Einstein's less famous argument for incompleteness of quantum mechanics}

In June of 1935 Einstein wrote  to Schr\"odinger \cite{einstein} bemoaning that the EPR paper \cite{epr} `buried in the erudition' the simplicity of the point he was trying to make \cite{howard}. In this letter he defines \emph{completeness} of a state description as
\begin{quotation}
\ldots$\Psi$ is correlated one-to-one with the real state of the real system\ldots
\end{quotation}
and a \emph{separation hypothesis} between systems enclosed in different boxes as
\begin{quotation}
\ldots the second box, along with everything having to do with its contents, is independent with regards to what happens to the first box (separated partial systems)\ldots.
\end{quotation}
For our purposes it is only important that the separation hypothesis, together with the assumption of `real states of real systems' implies local realism, although it potentially encompasses more.

Einstein goes on to consider entangled particles A and B, and to point out that depending on the choice of \emph{kind of measurement} on A (the type of observable, not its outcome) we ascribe different state functions $\Psi_B$, $\underline{\Psi_B}$ to system B.
\begin{quotation}
The real state of B thus cannot depend upon the kind of measurement I carry out on A. (``Separation hypothesis'' from above.)
But then for the same [real] state of B there are two (in general arbitrarily many) equally justified [quantum states] $\Psi_B$, which contradicts the hypothesis of a one-to-one or complete description of the real states.
\end{quotation}

Einstein's description does not, in fact, carefully distinguish the ensemble of quantum states which are obtained on B for a single fixed measurement on A and the different ensembles of quantum states which correspond to distinct choices of measurement on A \cite{einsteinargument}. In a moment we will emphasize why the choice of different (and incompatible) measurements on A was a necessary part of his argument (and amusingly he points out, contra-EPR, that he `doesn't give a damn' whether the states $\Psi_{B}$, $\underline{\Psi_B}$ are eigenfunctions of observables on B!), but first let's note that this description of all the possible ensembles achievable in such experiments was subsequently carefully characterized by Schr\"odinger who, within a year, proved the quantum steering theorem \cite{ES,steeringnote}:\\
\begin{thm} Given an entangled state $|\psi_{AB}\>$ of two
systems $A,B$, a measurement on system $A$ can collapse system $B$
to the ensemble of states $\{|\phi_i\>\}$ with associated probabilities
$p_i$, if and only if
\[
\rho_B=\sum_i p_i |\phi_i\>\<\phi_i|,
\]
where $\rho_B\equiv Tr_A|\psi_{AB}\>\<\psi_{AB}|$ is the reduced
state of system $B$.
\end{thm}

The reason two (incompatible) measurements are necessary for Einstein's argument for incompleteness is that if one considers only a single measurement on A it is trivially possible to maintain a one-to-one correspondence between a real state $\lambda$ of system B and the quantum state: in this setting, the steering statistics for an entangled state on AB is indistinguishable from those of a mixture of quantum/real states $\{|\phi_i\>\leftrightarrow \lambda_i\}$ for B, arranged such that the measurement on A needs only reveal only which member of the ensemble pertains. By choosing to steer to one of two different ensembles of quantum states $\{\ket{\phi_i}\}$, $\{\ket{\phi_i'}\}$ \emph{with at least some elements distinct} this is no longer possible.

Einstein concluded that, assuming local realism, many different quantum states must be associated with any given real state of B. Note, however, that since these different quantum states for B are operationally distinct, it clearly cannot be the case that those different quantum states are all only ever associated with that one single real state of B. They must somehow differ in the ensemble of real states they correspond to. Such a difference can be reflected either in terms of the members of the ensemble (i.e. sometimes being associated with completely different real states) or in terms of the frequencies (probabilities) over the ensemble, or both.

In the EPR paper the initial state of AB used is maximally entangled, and the ensembles steered to are those of orthogonal quantum states (position or momentum eigenstates). For this steering scenario it is well known (see e.g. \cite{BRS}) that the Wigner function provides a local (but as per Einstein's argument, necessarily incomplete) description of reality.  We begin by showing that steering between two ensembles of orthogonal states for a qubit also can be explained within such a local realistic theory.

\section{Steering between 2 ensembles of orthogonal states}\label{2steer}

Let us formalize Einstein's conclusion and its implications, simplifying to the easiest case possible:  two different measurements on A that steer the quantum state of a qubit B to ensembles $\{|x\>,|X\>\}$, $\{|y\>,|Y\>\}$ where $|x\>,|X\>,|y\>,|Y\>$ are all different, $\<x|X\>=0=\<y|Y\>$, and the members of each ensemble are equally likely. As Schr\"{o}dinger had proven, this is possible for any entangled state $\ket{\psi_{AB}}$ for which 
\be
\rho_B=\hh \ket{x}\bra{x}+\hh \ket{X}\bra{X}=\hh \ket{y}\bra{y}+\hh \ket{Y}\bra{Y}.
\ee
In a realistic description, every quantum state corresponds to a probability distribution over a set of real states $\lambda$. When an entangled quantum state is prepared on AB, let $\nu(\lambda)$ denote the ensemble of real states for B. It must be the case that $\nu(\lambda)$ can be resolved into the steering ensembles as
\be\label{nu}
\nu(\lambda)=\hh x(\lambda)+\hh X(\lambda)=\hh y(\lambda)+\hh Y(\lambda)
\ee
where $\mu(\lambda)$, $\mu=x,X,y,Y$ denotes the probability density over real states corresponding to the quantum state $|\mu\>$. Einstein's argument then runs that while $x(\lambda),X(\lambda)$ could potentially have disjoint support, thereby still allowing for the possibility each $\lambda$ is associated with a unique quantum state, the incompleteness of quantum theory is assured by the fact that at a given $\lambda$ for which (say) $x(\lambda)$ is non-zero, one or other of $y(\lambda),Y(\lambda)$ must be non-zero. 

Now, Einstein (explicitly) and Schr\"odinger (at least for the sake of argument) assumed that a complete description of reality is possible, and that in such a theory the quantum state would therefore be incomplete in the precise sense Einstein defined \cite{ESincomplete}. Even for the simple case of steering  between 2 ensembles captured by the generic decomposition of equation \eqref{nu} this yields some extra consistency conditions which need to be satisfied. For example it must be possible to find a probability density $\nu(\lambda)$ over some space of real states that can be decomposed into probability densities $x(\lambda),X(\lambda)$ which are disjoint, because $|x\>$ and $|X\>$ are orthogonal. Denoting by $S_\mu$ the support of the probability density $\mu(\lambda)$ we have that
\be
S_\nu=S_x\cup S_X=S_y \cup S_Y.
\ee

In the original EPR argument the scenario considered involves steering of B  between the ensembles of position and momentum, and then analysis of the conclusions that can be drawn if a subsequent position/momentum measurement is performed on B. Similarly here we analyse the restrictions that the incomplete description of reality must obey if measurements of the projectors onto the ensemble - i.e. $\{|x\>\<x|,|X\>\<X|\}$ or $\{|y\>\<y|,|Y\>\<Y|\}$ are performed. Such consideration shows that we must also obey consistency conditions of the form
\be\label{int}
\int_{S_{x}}\!\!d\lambda\;y(\lambda)=|\<x|y\>|^2\equiv\alpha,
\ee
to conform with the probability of obtaining the outcome $|x\>\<x|$ if a measurement in the basis $|x\>,|X\>$ is performed on B after the quantum state has been steered to $|y\>$.

It is useful to identify 4 disjoint regions of the space of real states: $S_1\equiv S_x \cap S_y$, $S_2\equiv S_x \cap S_Y$,
$S_3\equiv S_X
\cap S_y$, $S_4\equiv S_X \cap S_Y$ and to use the notation
\be\label{notation}
\mu_j\equiv \int_{S_{j}}\!\!d\lambda\;\mu(\lambda),\;\;\; j=1,\ldots,4.
\ee
Since $|\<x|y\>|^2=|\<X|Y\>|^2=\alpha$ and $|\<x|Y\>|^2=|\<X|y\>|^2=1-\alpha$, from equations of the form (\ref{int}) we must have:
\begin{eqnarray*}
&x_1=y_1=X_4=Y_4=\alpha& \\
&x_2=y_3=X_3=Y_2=1-\alpha&
\end{eqnarray*}
with all other values 0 or 1.

By integrating (\ref{nu}) over the appropriate regions we identify a final set of consistency conditions:
\be\label{12}
\nu_j = \hh x_j+\hh X_j=\hh y_j+\hh Y_j, \;\;\; j=1,\ldots,4.
\ee
These are satisfied by taking $\nu_1=\nu_4=\alpha/2$, while $\nu_2=\nu_3=(1-\alpha)/2$.

So far then, all of these essentially trivial consistency conditions - which any local incomplete description of reality must obey - are easily complied with.

In Section \ref{3steer} we will show that if we add the possibility of an extra measurement on A being used to steer to a third ensemble of orthogonal states we find a contradiction, indicating any realistic theory explaining quantum theory must be nonlocal - Bell's theorem. However, we now turn to a proof that yields the same conclusion, but uses steering between only two ensembles, one of which contains a pair of non-orthogonal states.

\section{Steering between 2 ensembles, one of which contains non-orthogonal states, implies the untenability of local realism}\label{2steernon}

\begin{figure}[t!]
\begin{center}
\includegraphics[trim=1cm 6cm 1cm 6cm, width=2.5in]{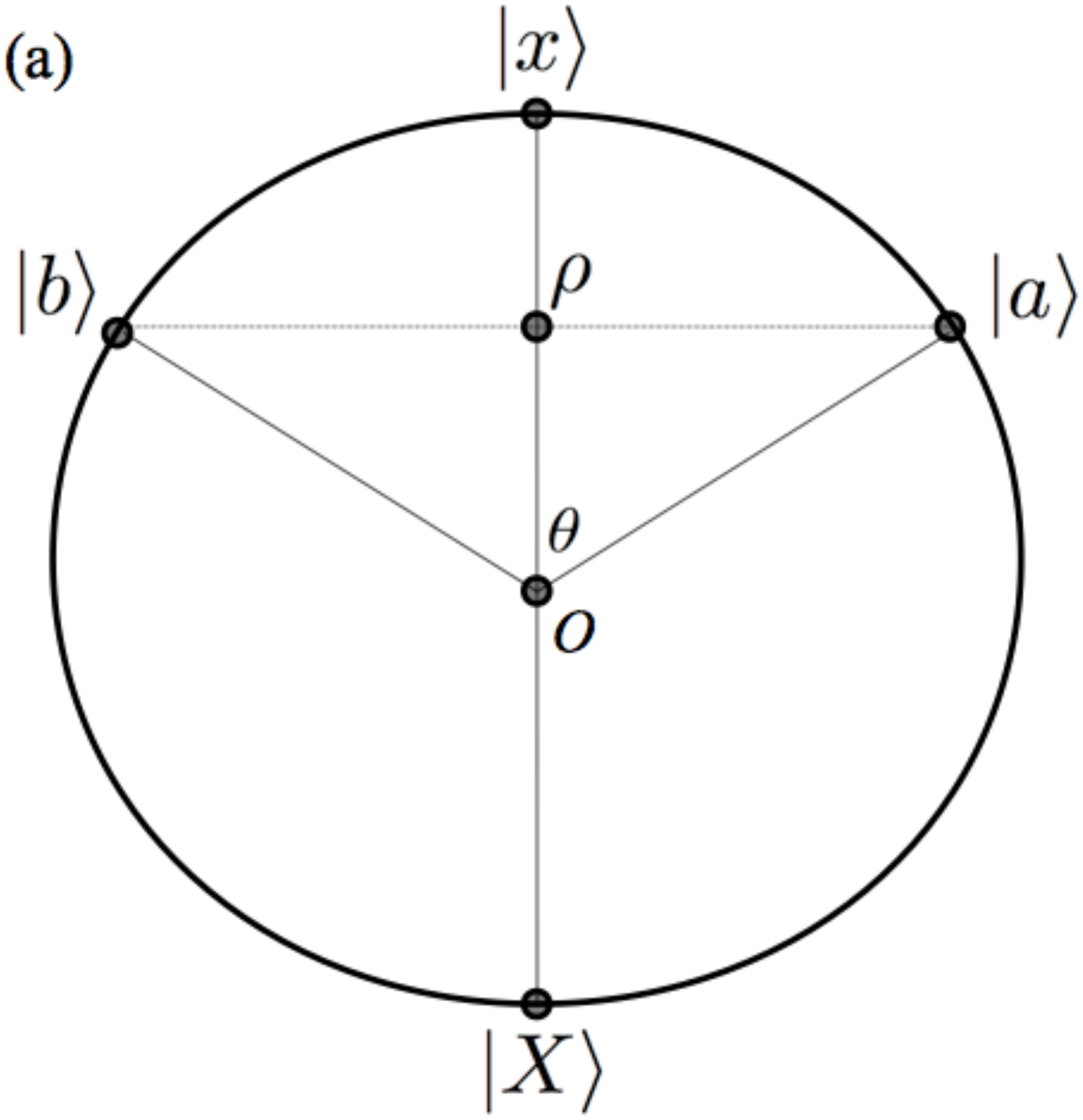}
\includegraphics[trim=1cm 6cm 1cm 6cm, width=2.5in]{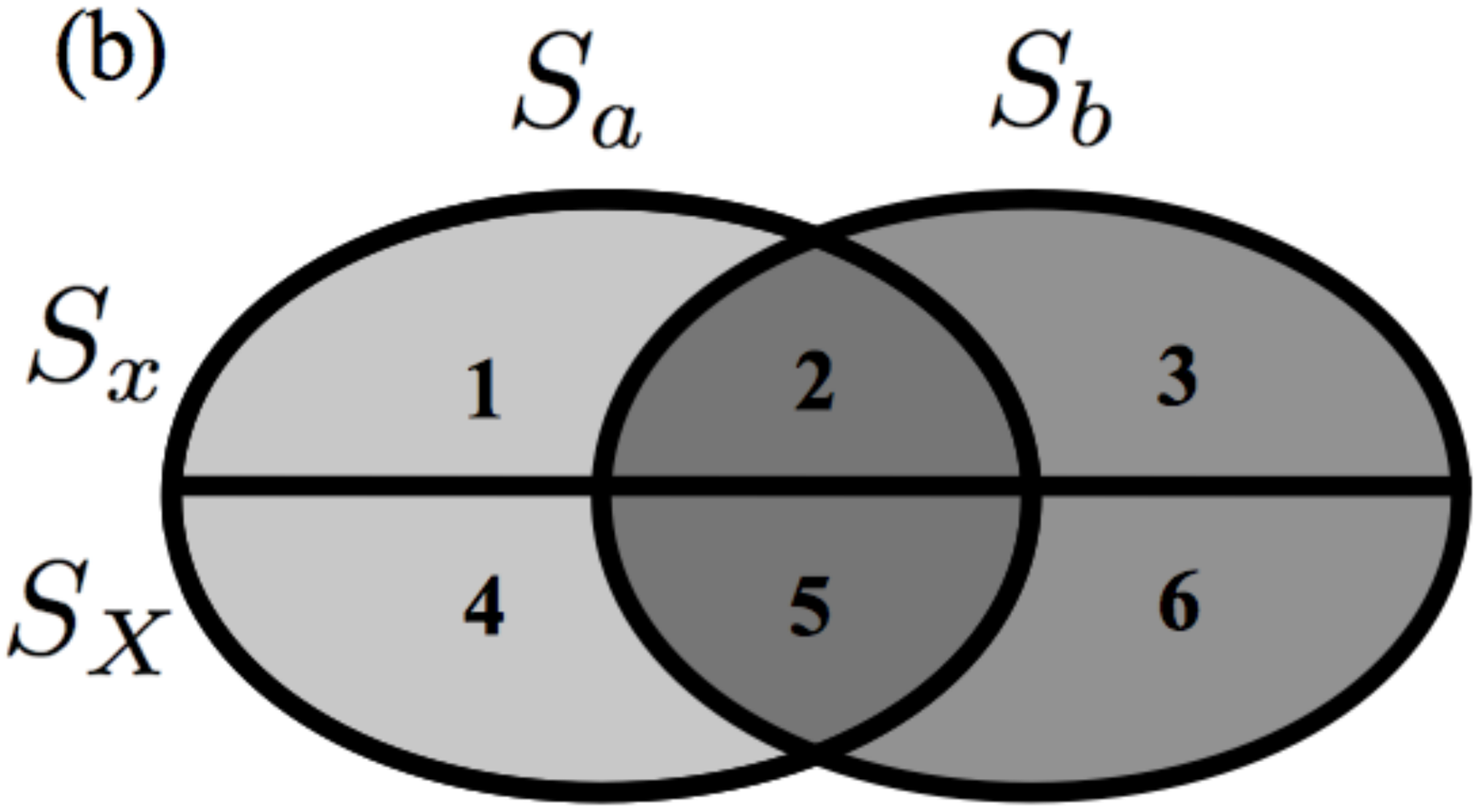}
\caption{(a) Two ensembles of $\rho$. (b) The space of real states, with disjoint regions of support labelled 1,..,6 such that $S_x = S_1\cup S_2\cup S_3$, $S_X = S_4\cup S_5\cup S_6$, $S_a = S_1\cup S_2\cup S_4\cup S_5$, $S_b = S_2\cup S_3\cup S_5 \cup S_6$. 
}
\label{xXab_steer}
\end{center}
\end{figure}

To show that incompleteness cannot save local realism we now consider the possibility of steering a qubit between two ensembles, where one of the ensembles contains non-orthogonal states. Most probably Einstein, but certainly Schr\"odinger, knew that this was possible - it is consistent with Einstein's calculation summarized in footnote \cite{einsteinargument} and is mentioned explicitly in Schr\"odinger's proof of the steering theorem (he limits only to ensembles wherein the states are linearly independent).

Consider then steering the state $\rho$ depicted in Fig.~1(a) either to its eigen-ensemble $\{\ket x,\ket X\}$ or to the ensemble $\{\ket a,\ket b\}$ that is an equal mixture of non-orthogonal states $\ket a = \cos\frac\theta 2 \ket x +  \sin\frac\theta 2 \ket X$ and $\ket b = \cos\frac\theta 2 \ket x -  \sin\frac\theta 2 \ket X$, where $\theta \in (0, \frac\pi 2)$. In the incomplete theory
these quantum states correspond to preparation of real physical states according to probability distributions $\mu(\lambda)$, $\mu = x,X,a,b$ which have support on sets labelled $S_\mu$. Local realism implies $S_x \cup S_X=S_a \cup S_b$; moreover because $\ket x$ and $\ket X$ are orthogonal, $S_x \cap S_X = \emptyset,$ while $S_a$ and $S_b$ can overlap. This is depicted in Fig.~1(b) where a convenient labelling for various regions of support is shown.

Once again using the notation in equation \eqref{notation} we have some simple consistency conditions, for example normalization imposes
\be\label{normX}
X_4+X_5+X_6=1.
\ee
Consider the case the system has been steered to the quantum state $|X\>$ and we wish to compute the probability of obtaining a measurement outcome $|a\>\<a|$ or alternatively of $|b\>\<b|$. A quick glance at Fig.~1 leads naturally to the constraints 
\begin{eqnarray}\label{Xab}
X_4+X_5&=&|\<a|X\>|^2=\sin^2\frac{\theta}{2} \nonumber\\
X_5+X_6&=&|\<b|X\>|^2=\sin^2\frac{\theta}{2},
\end{eqnarray}
and for $\theta\in(0,\pi/2)$ these are readily seen as incompatible with Eq.~(\ref{normX}). For example summing these two equations and inserting the normalization constraint implies $1+X_5=2\sin^2\theta/2=1-\cos\theta$, implying $X_5<0$. As $X_5$ is the integral of a probability density it must be positive, in the face of this it is clear that the assumption of local realism, upon which the whole discussion is premised, must be false - this is Bell's Theorem.

\subsection{The subtlety of \emph{deficiency}}

The steering-based proof of Bell's theorem we have just presented has actually made a subtle assumption. Let us reconsider the case wherein the system has been steered to the quantum state $|X\>$ and we wish to compute the probability of obtaining a measurement outcome $|a\>\<a|$. Above we assumed this imposes $X_4+X_5=|\<a|X\>|^2=\sin^2\theta/2$. However, while it is certainly the case that all real states $\lambda$ in $S_a$ must deterministically yield the outcome $|a\>\<a|$, any real states in $S_b$ can potentially also yield this outcome, because $\ket a$ and $\ket b$ are not orthogonal. Within regions 3 and 6 they need not even do so deterministically, because $0<|\<a|b\>|<1$. That is, the incomplete realistic theory may have a property defined in \cite{harriganrudolph} as \emph{deficiency}. In fact, for the case of three and higher dimensional quantum systems, it follows \cite{harriganrudolph} from the Kochen-Specker theorem that any realistic theory (local or otherwise) \emph{must} be deficient, that is, it must be the case that the set of real states that a system, prepared in quantum state $|a\>$, may actually be in, is necessarily strictly smaller than the set of real states which would reveal the measurement outcome $|a\>\<a|$ with some finite probability.

To see that allowing the local realistic theory to be deficient does not save it from Bell's Theorem we formally capture the possibility of deficiency by defining a \emph{response function} or \emph{indicator function} $\xi_a(\lambda)$ , which for every $\lambda$ is simply the probability that particular real state yields the $|a\>\<a|$ outcome. Since $0\le\xi_a(\lambda)\le1$ we must have
\be
0\le {X_6^{(a)}}\equiv\int_{S_{6}}\!\!d\lambda\;\xi_a(\lambda)X(\lambda)\le\int_{S_{6}}\!\!d\lambda\;X(\lambda)=X_6,
\ee
and so we obtain generalizations of Eq.~\eqref{Xab}
\begin{eqnarray}\label{Xab2}
X_4+X_5+{X_6^{(a)}}=\sin^2\frac\theta2 \nonumber\\
{X_4^{(b)}}+X_5+X_6=\sin^2\frac\theta2.
\end{eqnarray}

Once again equations \eqref{normX},\eqref{Xab2} cannot be simultaneously satisfied. For example, substituting \eqref{normX} into \eqref{Xab2} gives
\bea
X_6&=& {X_6^{(a)}}+\cos^2\frac\theta2 \\
X_4&=& {X_4^{(b)}}+\cos^2\frac\theta2
\eea
and so
\be
X_4+X_6\ge2\cos^2\frac\theta2>1, \text{ for }\theta\in(0,\pi/2),
\ee
which violates normalisation.

\section{Steering between 3 ensembles of orthogonal states}\label{3steer}

We now return to the steering of Section \ref{2steer}, and show that steering between 3 ensembles of orthogonal states can sometimes violate the consistency conditions. For simplicity presume for the moment that the third ensemble of orthogonal states $\{\ket z, \ket Z\}$ is such that the state $|z\>=(|x\>+|y\>)/\sqrt{2(1+\alpha)}$, i.e.  $\ket z$ `bisects' the states $|x\>,|y\>$. Then
\[
|\<z|x\>|^2=|\<z|y\>|^2=|\<Z|X\>|^2=|\<Z|Y\>|^2
=\frac{1+\sqrt{\alpha}}{2}\equiv\beta,
\]
the last term being the quantum mechanical Born-rule prediction. For the same regions of support $S_i$ defined in Section \ref{2steer} we now deduce consistency conditions
\bea
z_1+z_2=z_1+z_3=Z_3+Z_4=Z_2+Z_4&=&\beta\\
z_3+z_4=z_2+z_4=Z_1+Z_2=Z_1+Z_3&=&1-\beta
\eea
Clearly $z_2=z_3$ and $Z_2=Z_3$. We must also have
\be
\nu_j=\hh z_j+\hh Z_j \;\;\; j=1,\ldots,4.
\ee

There is no way to satisfy all these equations, subject to the
requirement $z_j,Z_j\ge 0$. For example, an independent
set of the above equations is
\bea
z_1+z_2=Z_2+Z_4&=&\beta \label{aa}\\
z_2+z_4=Z_1+Z_2&=&1-\beta \label{bb}\\
z_1+Z_1&=&\alpha \label{cc}
\eea
From these we obtain
\begin{eqnarray*}
Z_1&=&\alpha-z_1=\alpha-(\beta-z_2)=\alpha-\beta+(1-\beta-z_4)\\
&=&1-2\beta+\alpha-z_4,
\end{eqnarray*}
which, using $\beta=\hh(1+\sqrt{\alpha})$, gives
$Z_1=\alpha-\sqrt{\alpha}-z_4$. This is manifestly negative for
any $0< \alpha,z_4<1$.

Once again, the failure to keep the incomplete realistic models consistent indicates the initial assumption of local realism is unviable.

It is interesting to note that if the states $\{|x\>,|y\>,|z\>\}$ had been chosen to be the eigenstates of the Pauli operators $\sigma_x,\sigma_y,\sigma_z$ then $\alpha=\beta=1/2$ and the consistency conditions \emph{would} be satisfiable. Contrast this with the fact that the inconsistency obtained when $|z\>$ bisects $|x\>,|y\>$ holds regardless of how close these latter two (distinct) states are. As such we see that ``how far apart'' the triples of states are does not capture the difficulty or otherwise of reproducing their steering properties in an incomplete model of physical reality.

To investigate this further we have analyzed the possible triples of overlaps
\be
\alpha\equiv|\<x|y\>|^2, \beta\equiv|\<x|z\>|^2, \gamma\equiv|\<y|z\>|^2
\ee
which do or do not allow for a proof of the untenability of local realism by violating (or otherwise) the consistency conditions above. A mix of analytical and numerical evidence makes us confident that the answer takes the particularly pleasing form given by
\begin{conjecture}
The triple of overlaps $\alpha,\beta,\gamma$ demonstrate a violation of local realism if and only if the point
$[\alpha,\beta,\gamma]$ does not lie in the convex hull of the four points $[1,0,0]$, $[0,1,0]$, $[0,0,1]$, $[1,1,1]$.
\end{conjecture}

We have also performed some preliminary forays into the question of how steering between four ensembles may differ. One thing we noticed in this regard is that if we look at steering performed on a Werner state (mixture of maximally entangled and maximally mixed state) then the lowest probability of the maximally entangled state for which violation of local realism can still be demonstrated is 4/5 when steering between 3 ensembles and $1/\sqrt{2}$ when steering between 4 ensembles.

\section{Outlook}
In deriving Bell's Theorem from steering, a number of observations crop up that merit further investigation.

One intriguing feature is that we never make use of the assumption $\mu(\lambda)\ge0$, rather positivity was required only for integrals of the distributions over certain regions within the space of real states. Thus the proof rules out certain options for quasi-representations of the quantum state as well.

A second observation concerns the deficiency property introduced in Section \ref{2steernon}. It has been shown \cite{harriganrudolph} that measurement-outcome contextuality \cite{kochenspecker} manifests itself in incomplete models of reality via \emph{deficiency}. This in turn makes it strictly impossible for such a model to obey conditions of the form in equation \eqref{int} - i.e. all the non-orthogonality of quantum states cannot be attributed to classical non-orthogonality of their associated probability distributions. Perhaps combining this observation with steering of entangled systems of dimension three or higher can yield different steering-type proofs that local realism is untenable.

Finally, the proof in Section \ref{2steernon} did not require an equation of the form 
\be
\nu(\lambda)=p x(\lambda)+ (1-p) X(\lambda)=\hh a(\lambda)+\hh b(\lambda)
\ee
to hold. Not only did the probabilities with which elements of the ensemble appear play no role, the proof would have still gone through even if \be
p x(\lambda)+ (1-p) X(\lambda)\neq \hh a(\lambda)+\hh b(\lambda),
\ee
as long as the supports of the distributions satisfied $S_x\cup S_X=S_a\cup S_b$. Thus only a weaker assumption than \emph{preparation non-contextuality} as defined by Spekkens \cite{spekkens} is needed. It may be interesting therefore to consider further a weaker version of preparation contextuality, one defined solely in terms of the equivalence, or otherwise, of the supports of distributions which convexly combine to the same mixed state, and not an exact equivalence of the convexly combined probability densities themselves. In this regard we should mention that Barrett has shown \cite{barrett} that standard bipartite proofs of Bell's theorem can be converted into a proof of preparation contextuality.


\section{Conclusions}

Before concluding let us mention some relevant work. The proof in Section \ref{2steernon} is readily extended to show nonlocality for all non-maximally entangled pure states, reproducing the conclusions of Gisin, Popescu and Rohrlich \cite{gisin}. Although our proofs are algebraic and thus reminiscent of GHZ \cite{GHZ} and Hardy \cite{hardy} type arguments against local realism, the proof in Section \ref{3steer} would seem closest to Mermin's exposition of Bell inequalities in \cite{mermin}. Harrigan and Spekkens \cite{nicnic} perform a more careful and thorough exposition of Einstein's argument above for incompleteness and the relationship to locality. In \cite{werner}, Werner presents an alternative route which could have led Einstein to Bell's argument.
 

Finally, one may wonder whether the quantum state can still be argued to be incomplete in Einstein's sense above even when separability is given up. While it is in fact possible to obey all consistency conditions generalizing those above for such an incomplete realistic theory \cite{lewis}, it turns out that an assumption of separability for \emph{product} quantum states leads to the exact opposite conclusion, namely that the quantum state must be complete \cite{PBR}.

In conclusion, if Einstein and Schr\"odinger had probed only a little further into whether an incomplete description of physical reality can actually fully explain the gedankenexperiment that they had used to rule out completeness of quantum theory, perhaps the tension between locality, realism and quantum theory would have been brought to the fore significantly earlier.

\acknowledgements
TR acknowledges multiple conversations with, and insights provided by, Rob Spekkens, both with regards to steering as both a practical tool in quantum cryptography and as a foundational foil. We also appreciated useful comments by Matt Pusey.  We are indebted to  Guido Bacciagaluppi and and Elise Crull  for providing us in advance of publication their complete translation of the letter from Einstein to Schr\"odinger and to Don Howard for providing a copy of the original. TR supported by the Leverhulme trust. SJ is funded by EPSRC grant EP/K022512/1.



\begin{thebibliography}{99}
\bibitem{einsteinargument} To be more concrete, paraphrasing the translation of the letter in \cite{guido}, Einstein says consider an entangled state expanded in two ways:
\[
\psi_{AB}=\sum_{mn}c_{mn}\psi_m(x_1)\chi_n(x_2)=\sum_{mn}\underline{c_{mn}}\underline{\psi_m(x_1)}\chi_n(x_2),
\]
where $\psi(x_1)$, $\underline{\psi(x_1)}$, $\chi(x_2)$ are ``eigen-$\psi$''  of commuting systems of observables $\alpha,\underline{\alpha},\beta$. If one makes an $\alpha$ (resp. $\underline{\alpha}$) measurement on A the state on B reduces to ${\psi_B}=\sum_{n}c_{mn}\chi_n(x_2)$ (resp. $\underline{\psi_B}=\sum_{n}\underline{c_{mn}}\chi_n(x_2)$) and all that is required for us to conclude that the ``$\psi$-description'' is not in one-to-one correspondence to the real state is that ${\psi_B}$, $\underline{\psi_B}$ are ``at all different from each other''.
\bibitem{ESincomplete} Schr\"odinger considered that certain quantum states - entangled states or macroscopic superpositions were unattainable, which may have nullified Einstein's specific argumnet for incompleteness. However given Schr\"odinger's concerns about both quantum jumps and the probabilistic nature of quantum predictions it is probably safe to say that he would still have assumed the theory incomplete in some sense even if entangled states were somehow excised.
\end{thebibliography}
\end{document}